\documentclass[oribibl]{llncs}
\usepackage{graphicx}
\usepackage{subfig}
\usepackage{booktabs}
\usepackage{epstopdf}
\usepackage{tikz}
\usepackage{pgf-umlsd}
\usepackage{booktabs} 

\usepackage{threeparttable}
\usepackage{multirow}

\usepackage{color}
\newcommand*{\circled}[1]{\lower.7ex\hbox{\tikz\draw (0pt, 0pt)%
    circle (.5em) node {\makebox[0.7em][c]{\small #1}};}}

\begin{document}

\title{Typer vs. CAPTCHA: Private information based CAPTCHA to defend against crowdsourcing human cheating}
 

\author{Jianyi Zhang\inst{1}\and
Xiali Hei\inst{2} \and
Zhiqiang Wang\inst{1,3}}

\institute{Beijing Electronics Science and Technology Institute, China \\
\email{zjy@besti.edu.cn}\\
University of Louisiana at Lafayette,USA \\
\email{xiali.hei@louisiana.edu}\\
}

\maketitle              

\begin{abstract} 
Crowdsourcing human-solving or online typing attacks are destructive problems. However, studies into these topics have been limited. In this paper, we focus on this kind of attacks whereby all the CAPTCHAs can be simply broken because of its design purpose. After pursuing a comprehensive analysis of the Typer phenomenon and the attacking mechanism of CAPTCHA, we present a new CAPTCHA design principle to distinguish human (Typer) from human (user). The core idea is that the challenge process of the CAPTCHA should contain the unique information with a private attribute. The notion of our idea is based on the information asymmetry between humans. Without this private information, Typers will not be able to finish the attack even if they recognize all the characters from the CAPTCHA. 

We formalize, design and implement two examples on our proposed principle, a character-based, and a datagram-based case, according to a web interaction and password handling program. We challenge the user to select the password from the random characters that are not in the password sequence or to place the randomly sorted sequences into the correct order. A novel generation algorithm with a fuzzy matching method has been proposed to add the capability of human error tolerance and the difficulty of random guess attack. Unlike other solutions, our approach does not need to modify the primary authentication protocol, user interface, and experience of the typical web service. The several user studies' results indicate that our proposed method is both simple (can be solved by humans accurately within less than 20 seconds) and efficient (the Typer can only deploy a random guess attack with a very low success rate).
\end{abstract}

\section{Introduction}

Since the concept was first introduced by Von Ahn \cite{von2003captcha}, researchers develop a wide variety of CAPTCHAs (Completely Automated Public Turing test to tell Computers and Humans Apart). Nowadays, all techniques developed for generating CAPTCHA rely on the information that is known to the server but not to the user. The consequence of such information asymmetry is the user can only pass the Turing test by solving a hard AI problem. The information to be challenged thus is limited to random letters, such as Handwritten \cite{rusu2004handwritten} and reCaptcha \cite{von2008recaptcha}. Besides, the information can also be designed with high-entropy entries that need human’s cognitive or interactive ability, such as Asirra \cite{elson2007asirra}, ESP-PIX \cite{von2004labeling}, Google image orientation \cite{gossweiler2009s}, and Sketcha \cite{ross2010sketcha}. 

However, the design purpose of the CAPTCHA, easy for a human but hard for robots, has also been utilized by some parties to outsource the recognition piecemeal to the unskilled labor pools. The attackers can decode the CAPTCHA by simply relaying the recognition processes to human solvers (be referred to as the Typer). Unfortunately, to date, most of the existing methods cannot completely solve the Typer problem. 
In this paper, we present a new direction for avoiding crowdsourcing Typers cheating on CAPTCHA. We introduce the private information to get the asymmetry and break the same verification connection between the attackers and the Typers. 
Based on the commonly used password transmission process, we introduce a user study to demonstrate our principle, and it shares the common principle on user's identity verification. That is, the private information (i.e., the password) we embedded must come from the user's HTTP POST \cite{Post} to ensure that the verification process and the CAPTCHA challenge implements simultaneously.

Unlike other solutions, our approach does not need to modify the basic authentication protocol, user interface, or experience of the common web service. Also, the method covers most of the daily web application scenarios like registration or certificate identification. The results of several user studies and experiments indicate that our proposed method is simple and effective. No matter what kind of counter attacks (fake accounts, compromised accounts or password changing), the Typer can only deploy a random guess attack with a very low success rate, while a legitimate user can solve the CAPTCHA accurately within less than 10 seconds. 

We summarize our contributions as follows: 
\begin{itemize}
\item We conduct an in-depth study and analysis of the online CAPTCHA typing jobs and the Typer phenomenon.
\item We propose a new principle for prohibition of the Typers cheating. 
\item We create two demonstration examples of our method which can be utilized in the real scenarios. 
\item We devise a generation algorithm to add the capability of human error tolerance and the difficulty of random guess attacks. 
\end{itemize} 


\section{Background and Related Work}
\subsection{Human Solver and Online CAPTCHA Typing Jobs}
\begin{figure}[!t] 
\centering
\includegraphics[height=1.7in]{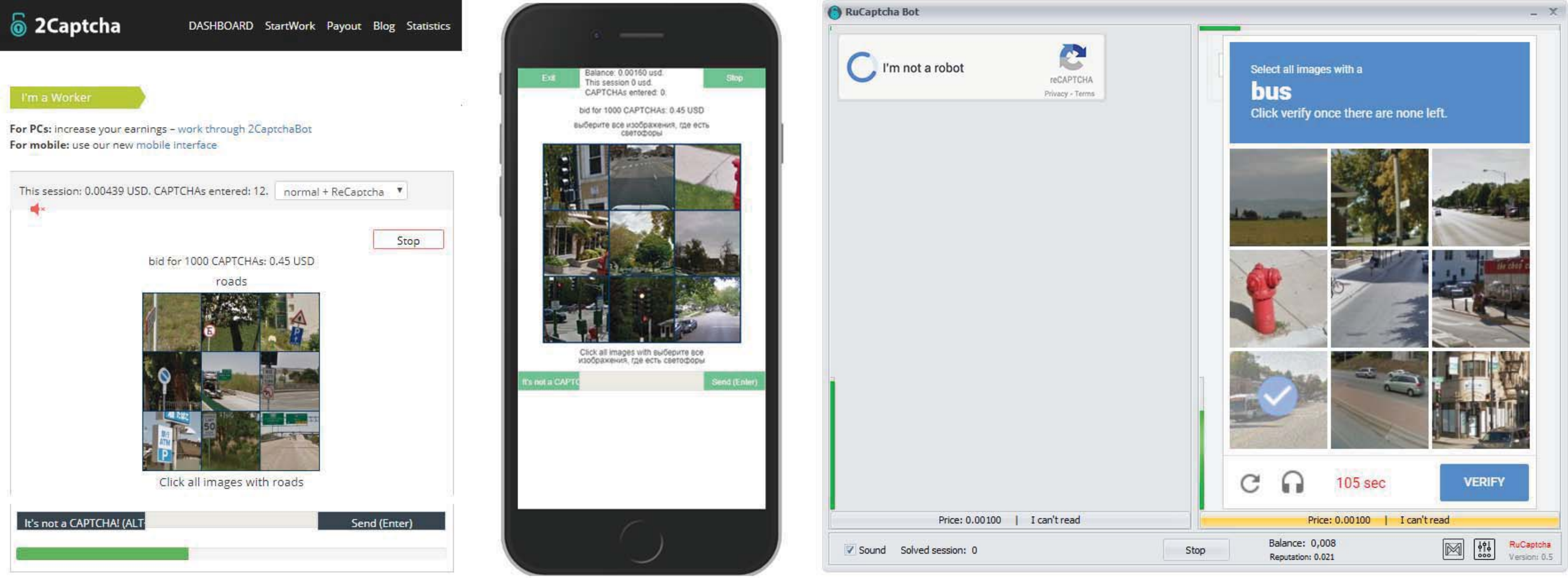}
\caption{The interface of the crowdsourcing CAPTCHA typing work. It contains web-based entry form or mobile version. And the right is the RuCaptcha software which the newest version has already supported the Google's reCaptcha \cite{von2008recaptcha}.}
\label{interface} 
\end{figure}

CAPTCHA should be simple and easy to be solved by humans on its original design. Therefore, their validation can be entirely sidestepped by outsourcing the task to human labor pools on a ``for hire" basis (or opportunistically \cite{von2003captcha}
). Nowadays, the emergence of new technologies, like reCaptcha \cite{von2008recaptcha}, deepCaptcha \cite{deepcaptcha}, and rtCaptcha \cite{rtcapt}, make it hard to automate the recognition of a CAPTCHA by robots. However, the CAPTCHA entry work, which only requires human labors' typing skills, is easy for the human labors who are referred as the CAPTCHA Typers. Figure \ref{interface} shows the interface of the Typers' entry work. Marti Motoyama \emph{et al.} have analyzed this problem in 2010 \cite{motoyama2010re}. Since then, such advertisements have increased significantly on ``work-for-hire" sites (labor market) such as \emph{upwork.com} and \emph{zbj.com}. After that, to automate the real-time solving of CAPTCHA, commercial malware industry setups a series of web services, which consist of the customer's front-end programs and the Typer's back-end programs.  

The customer's front-end programs, like \emph{2captcha.com}, \emph{De-Captcher.com} and \emph{Antigate.com}, are the CAPTCHA solving sites that the malware industry provides real-time CAPTCHA-to-text decoding services. 
And the Typer back end, like \emph{2captcha.com} (It both has the front end and back end service), \emph{Kolotibablo, Dama2} and \emph{PixProfit}, are the malicious crowdsourcing sites that pay the human labors to perform the CAPTCHA entry works. 

\begin{figure}[!t]
\centering
\includegraphics[height=2in]{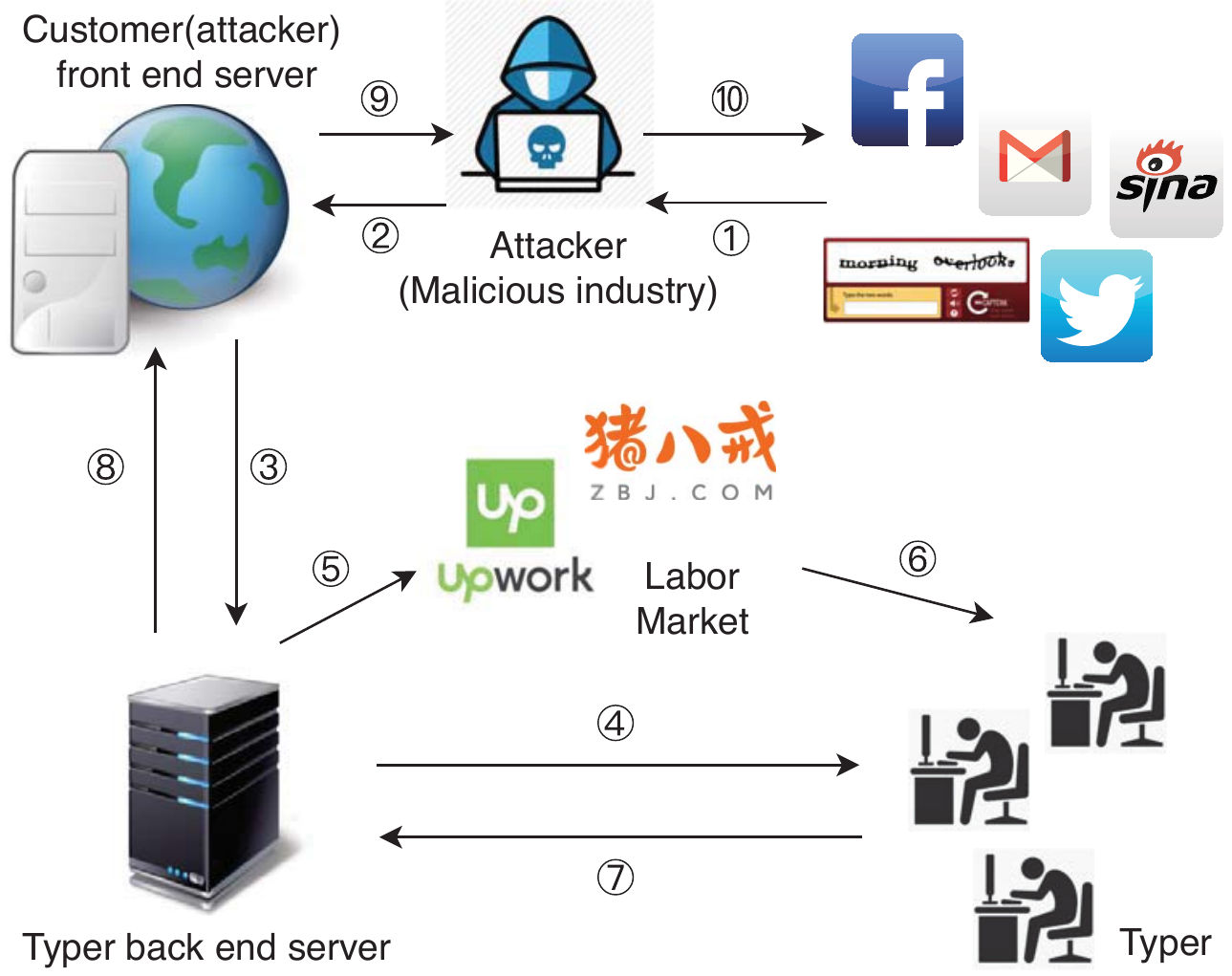}
\caption{Workflow of the online CAPTCHA typing jobs}
\label{typer}
\end{figure}

The example in Figure \ref{typer} illustrates a typical workflow of the CAPTCHA typing jobs and the relationships among the participants involved in the attacker and the Typer. Here we take the credential stuffing attack as an example. 
When the attacker attempts to hijack an existing account and accomplishes account takeover through automated web injection, a CAPTCHA will challenge them \circled{1}. After the attacker uploads the CAPTCHA to the customer (attacker) front-end API \circled{2}, the front end server queues the CAPTCHA to be solved on the (Typer) back end server \circled{3}. The back end then directly distributes the task \circled{4}, or in a form of a data-entry job on labor market sites\circled{5}\circled{6}, to a random Typer to obscure the login information. Then the Typer enters a result and the web service on the front end returns it to the API \circled{7}-\circled{9}. The attacker packages the account information with the recognition CAPTCHA texts. And in the end, the attacker finishes the web injection after sending the HTTP POST packages \circled{10}.
	
\subsection{Solutions to the Typer Problems}
The process of solving the Typer problem, also known as the ``relay attacks" or the ``3rd party human attack" problem, is to find a way of avoiding all crowdsourcing human cheating on CAPTCHA. Due to the simplicity, high success rate and low economic costs (usually less than one dollar per thousand CAPTCHAs), this problem is much more feasible in practice than machine vision or other automated attacks \cite{motoyama2010re}\cite{gentry2005secure}. The attacker only needs to simply forward the CAPTCHA to the human solvers, who type the challenge characters and wait for the Typers' response. Similarly, the video-based CAPTCHAs \cite{xu2012security} are straightforward as well by forwarding the video file or sending enough snapshots to the human solver. 

Unfortunately, to date, there is no particularly effective solution to overcome this problem. Here we distinguish three main techniques that focus on how to design new CAPTCHAs to avoid the Typer problem: One way is to set a ``per-character-response time" for solving the CAPTCHA. A detailed analysis of the reaction time suggests that the Typer can improve the success rate and the response speed through a series of training \cite{mohamed2014three}. Meanwhile, the most commonly encountered distortions of the CAPTCHA can also severely degrade human usability \cite{bursztein2010good}. 

Another approach is the user-interaction based CAPTCHA. It queries users to perform specific tasks, which are impossible for the artificial intelligence and relies on the stream interactions that a user performs while responses the CAPTCHA challenge. 
Ye \emph{et al.} have developed an improved method, named DDIM (Drag-n-Drop Interactive Masking), applies drag-n-drop and masking scheme to distinguish the legitimate user from the Typers and robots \cite{ye2013ddim}. Game CAPTCHA is an interesting example that challenges the user to execute a game-like cognitive action interacting with a series of images. Mohamed \emph{et al.} \cite{mohamed2014three,mohamed2014dynamic} have studied the performance of game-play behavioral features under the stream-based relay attacks which means the attacker can synchronously relay the data stream from the server over to a human solver, Typer, and then relay back with the response to the server. The results demonstrate that the Dynamic Cognitive Game (DCG) CAPTCHA offers some level of resistance to relay attacks. However, increasing the size of the game, the interaction between the user and the server not only improves the ability to resist the relay attack but also introduces complex operations that make users feel bored. And the simplicity of the underlying cognitive task cannot provide sufficient protection capability.

The use of information asymmetry is the third approach. It introduces unique information into in the CAPTCHA challenge process. 
This information can be purchase records, the people you familiar with, a geographical location \cite{wei2012geocaptcha}, Multiple Factors Authentication (MFA) information or different data that exists on the login page but outside the CAPTCHA itself \cite{halprin2009dependent}. But if the information does not have the private attribute, it can also be shared with the Typer because doing so is no threat to the account security. MFA, as it is commonly abbreviated, is a method of computer access control that adds extra channels to the basic authentication process. The most widely used channel of this form is the mobile phone. In this way, the clients can use a one-time-valid passcode, that is sent to user's mobile by SMS, in addition to their identities. This randomly generated passcode can be sent to user's mobile by SMS. Hence, the passcode can be seen as the CAPTCHA that the Typers cannot obtain directly. However, the emergence of shared line appearance (SLA) enables the attacker to configure a group of multiple mobiles that can each receive the SMS to a shared phone number. With the help of this technology, the malware industry has taken full advantage of the Online-SMS-receive services, like \emph{twilio.com, receive-sms-online.com} or \emph{shou-ma.com}, to receive the inbound messages that contain the verification code from applications' login requests, such as Facebook, Whatsapp, Yahoo, WeChat and so on. Thus, this method cannot completely solve the Typer problem. 
\section{Architecture and Methodology}
\subsection{Principle}
The key problem with the current way of preventing automated posting and login lies in the way how the tests are designed fundamentally. Today, all approaches are based on the Turing test of proving if a user is a human or bot. The main failure of this architecture is that the Turing test can only be used to distinguish between a conscious man and an unconscious machine. The test questions are randomly generated and independent of context. It cannot distinguish a human from another, that is, a Typer from a legitimate user. Having exploited this vulnerability, the attackers can bypass the CAPTCHAs via the Typers when the vision solving algorithms fail. 

To enable the CAPTCHA to distinguish a human (user) from another human (Typer), we should find the differences between them first. According to the analysis in Section II, the only way to solve this problem is to make the use of information asymmetry in a real-time scenario. The notion of our idea is simple but generic. If we can add unique information to the process, Typers will not be able to finish the recognition task if they do not know the data which is only shared by the user with the server. That is, we break the relationship between the attacker and the Typer with this unique information. By following this notion, we need to ensure that the attacker is not willing to share this information with the Typer or the attacker can not share the fake information with the Typer. Hence, the intuition behind our proposed method is that: \textit{\textbf{We embed the process of users' private data verification into the process of CAPTCHA challenge}}. 

Specifically, first, different users should share different information with the server. Second, the challenge processes should include and utilize this information. And third, for safety reasons, the challenge data should come from the users' submission instead of the servers' storage. We generate the CAPTCHA by attaching the private information that is submitted by the users. Meanwhile, the private attribute means they will not share the information with Typer. Afterwards, Typers cannot finish the job because they need to answer the questions with the private information included in the CAPTCHA.  
\subsection{Web Interaction Design and Case Study}
According to the commonly used processing method of password transmission, in the course of our work, we present an example algorithms as a case study to demonstrate our principle. The algorithm shares a common principle, and that is, we embed the private information (i.e., password, password hash or other privacy preserved information) into the Turing test to setup the information asymmetry, and this knowledge should come from the user's POST requests or submissions (cleartext, ciphertext or other forms) rather than from the system's database storage. 


\subsubsection{Case I: Character-based Example}
\begin{figure}[!t]
\centering
\subfloat{\includegraphics[height=2in]{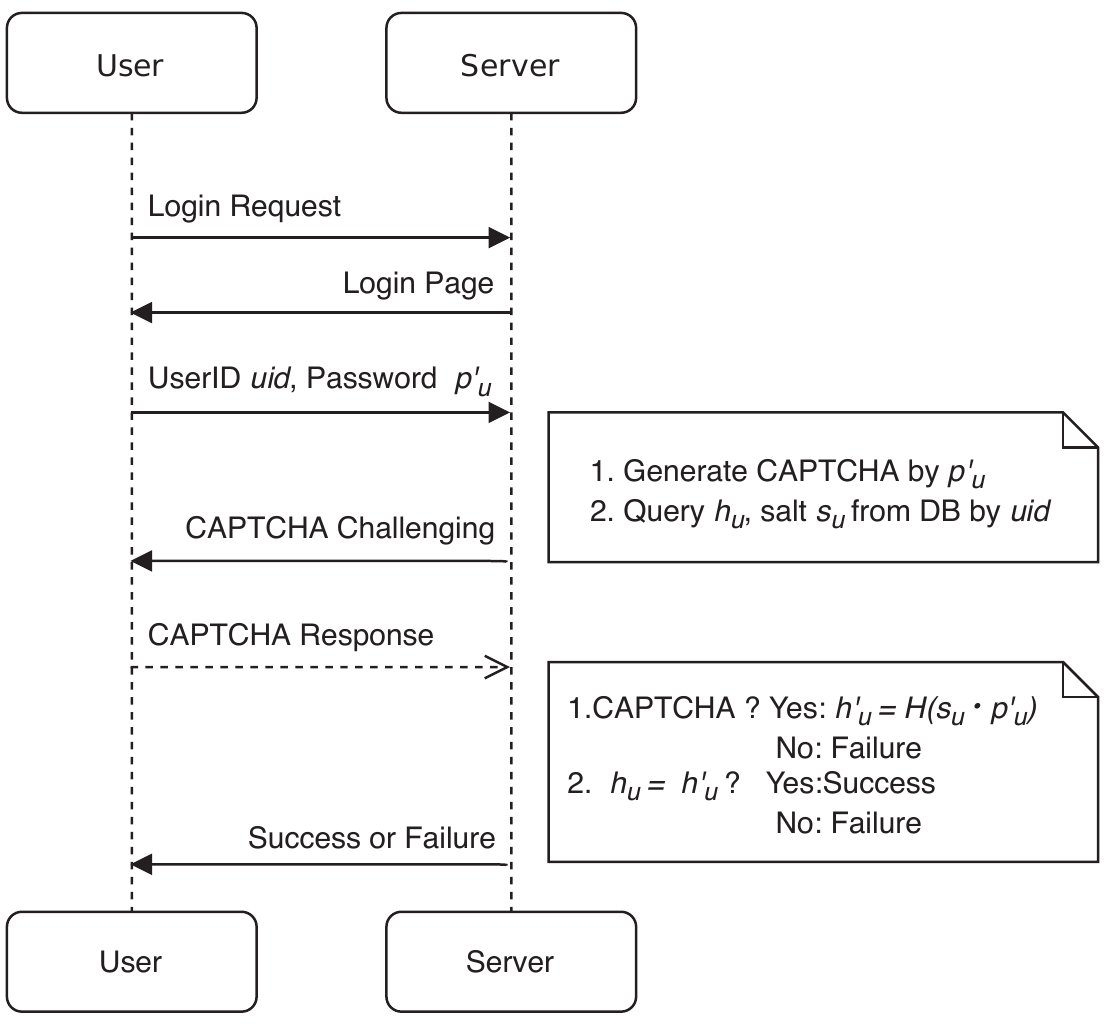}}
\hfil
\subfloat{\includegraphics[height=2in]{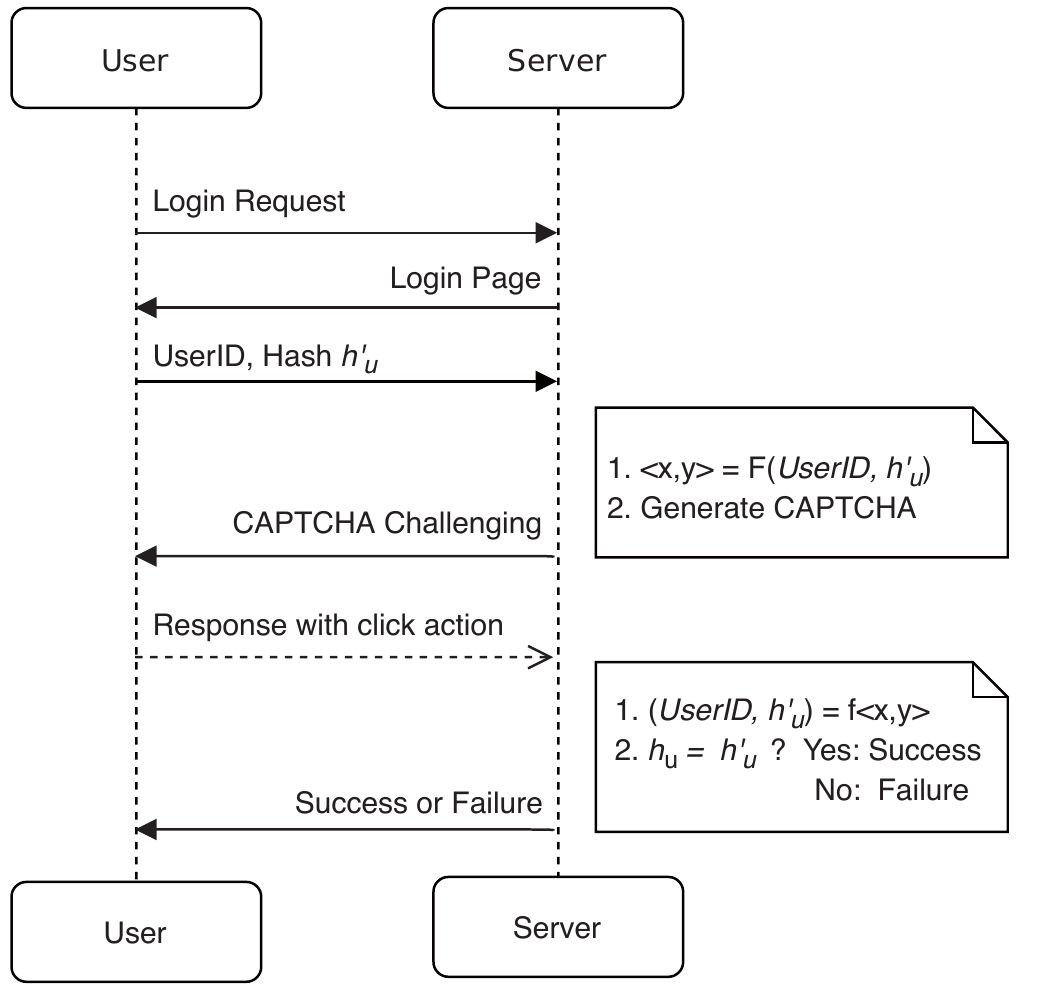}}
\caption{The sequence diagram of the authentication}
\label{CI+CII}
\end{figure}

According to the web interaction design, in most scenes, the webserver can get the plaintext content of the users' POST. Hence, the generation process of the CAPTCHA closely links to the character handling. For this case, a CAPTCHA is a sequence of random alphanumerical characters with the secret knowledge (password) provided by the user. A typical way to apply our proposed principle in user authentication is as follows. The webserver $WS$ stores a per-user salt ${s_u}$ mixed in with the salted-password hash ${h_u} = H({s_u} \cdot {p_u})$ for each UserID, where ${p_u}$ stands for the user's cleartext password. Upon receiving a login request, $WS$ sends the login page response to the client. Once the end-user transmits the UserID and password $p{'_u}$, the $WS$ generates the CAPTCHA by utilizing these credentials and sends the image to the user to click. After the user's clicks are recorded and sent, the $WS$ receives and analyzes these corresponding reactions. The user authentication continues only if these clicks are selected correctly. Otherwise, the webserver will return a failure result and interrupts the following process. Afterwards, if the CAPTCHA test is passed, the $WS$ calculates the hash value $h{'_u} = H({s_u} \cdot p{'_u})$ with the salt ${s_u}$ stored for the account, and compares the two hashes ${h_u}$ and $h{'_u}$ to determine whether they are the same. This process is called the character-based case. That is, the webserver can process every single character of the password that the user submits in the login request. The UserID and password $p{'_u}$ are attached to the CAPTCHA challenge processes of this case as the private information. Figure \ref{CI+CII} left shows the authentication sequence of the example.

The intuitive generation algorithm is to combine part of the password with random characters, and to ask users to select those that appear in the password. The $WS$ randomly and temporarily selects a certain number of letters from the cleartext of the password when it receives the user's credential. Meanwhile, it combines with random characters that are not in the password sequence which must not be clicked by the users.

\subsubsection{Case II, Datagram-based Example}
More generally, besides sending the cleartext password to the server through encrypted channels, to enhance the privacy, clients can also verify their identities by applying a challenge-response method to provide the secret information. Rather than merely asking for the password, the server also produces a value that depends on it. For example, the challenge-response protocols are designed by applying a one-way function to the passwords. On many systems, the one-way function is applied as a cryptographic hash and its output as a hashed password. Using hash-based challenge-response protocols along with TLS, obviously, will provide not only a secure session but also private protection between the user and server. However, the conflict between hash-based authentication and the capability of processing the shared secret from the user's POST makes us unable to use the character-based algorithm to generate the CAPTCHA. Hence, to extend the case I to a more general form, we employ the datagrams, based on the user's POST requests or information submissions, to generate the CAPTCHA. 

Taking the hash-based challenge-response authentication as an example, the workflow of our proposed approach in the user authentication is as follows. After receiving the login page, the end-user transmits the UserID and hashed password $h{'_u} = H(p{'_u})$ to $WS$, where $p{'_u}$ is the password that the user submitted. Then, the server calculates the UserID and $h{'_u}$ with $<x,y> = F(UserID, h{'_u})$ to cover the sequence of hash to CAPTCHA and sends the image to the user as the challenge. The processing of the Turing test is the same as the case I, and when the $WS$ receives the coordinates of the clicked point $<x,y>$, it then recovers the UserID and hash by function $(UserID, h{'_u}) = f<x,y>$ and compares the result with the hash value stored in the database. The CAPTCHA verification and the authentication succeed only if the two hash values are equal. The sequence of the datagram-based example is shown in Figure \ref{CI+CII} right.

There are many methods for the generation of the CAPTCHA by the user's POST datagram. In this paper, we use the simplest algorithm to demonstrate our principle. That is, we divide the string of the hashed password into different blocks and then shuffle them randomly to test the users whether they can drag the shuffled blocks in correct order. In our example, the length of the hash is 16bit, and we split it into six parts. For simplicity of this discussion, we have employed the click operation to place these sequences into a special order. 

\section{Generation Algorithm}
\subsection{Random Generation with Unpredictable} 
To increase the difficulty of random guess attacks and to introduce the level of uncertainties, our CAPTCHA generation algorithm leverages randomized approaches to assemble the CAPTCHAs. We randomly select the characters and then create a stochastic sequence to ensure the results are not predictable.

For the case of character-based selection, our approach divides the candidate subjects into two parts. It then internally labels them with the \emph{submission part} (SP) and the \emph{generation part} (GP). We select the SP from the private information shared between the server and the user (i.e., passwords). Because of the risk of password leaks and shoulder-surfing attacks, our method does not pick all the private information. Each time, we randomly choose some characters of the password to be the SP, and then randomly place them among the challenge characters. 
For the datagram-based case, our approach randomly splits the private information (i.e., password hash) into different parts and shuffles them into a random order. 
We then employ the clicking or dragging operation to be done by the users in the aim of sorting the random sequence into the correct order.
\subsection{Error Tolerance with High Reliability}
To increase the capability of human error tolerance and also add uncertainty to the process, we deploy a fuzzy matching method based on the users' responses. Since human beings may make mistakes even on the familiar things, we call these mistakes ``slight mistakes", we fully consider this flexibility of human judgment in the CAPTCHA solving task. 

In this paper, we consider a response to be positive if a user identities 3 out of 4 password characters correctly while selecting none of the GP. Simply treating these slight mistakes as correct may weaken the security of web services. Besides, if the Typer is lucky enough to pass the test with this slight error, they will use the characters they click in the following steps. Hence, we do not think twice slight mistakes are still insignificant. We deploy a fuzzy matching algorithm to take full advantage of evaluating the users' judgments. The partial credit algorithm (PCA) is a solution to improve the CAPTCHA which scores the responses more like a human \cite{elson2007asirra}. It considers two ``almost right" answers are strong evidence that the subject is a human. Our fuzzy matching method is built on the PCA with a similar scheme, but with the different attitude towards ``almost right" or slight mistakes. 
That is we build our hypothesis base on the fact that, a human may miss some objects from the the private knowledge but never select the characters that are not in the information.

\begin{figure}[t]   
\begin{minipage}[t]{0.45\linewidth}   
\centering   
\includegraphics[height=1.4in]{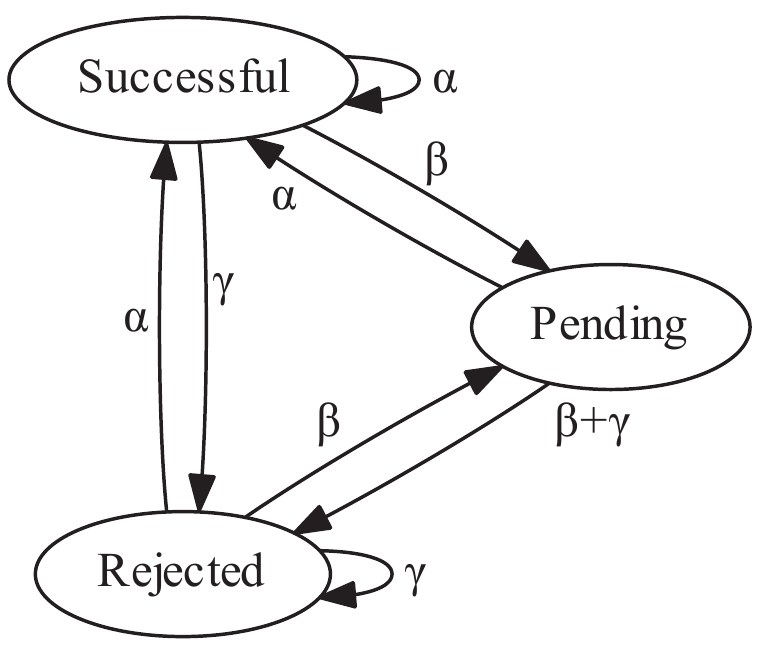}   
\caption{The FSM of the CAPTCHA generation process}   
\label{fsm}   
\end{minipage}   
\begin{minipage}[t]{0.2\linewidth} 
\centering
\end{minipage} 
\begin{minipage}[t]{0.5\linewidth} 
\centering   
\includegraphics[height=1.4in]{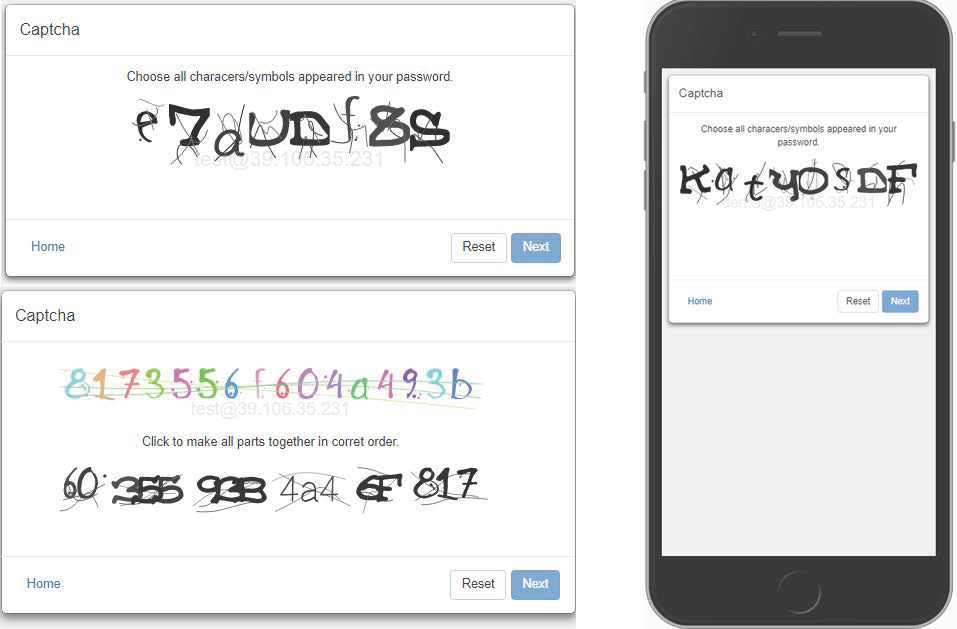}   
\caption{The user interface of our proposed method. 
} 
\label{impl3}   
\end{minipage}   
\end{figure}

Mathematically, we can state this relationship as follows. When the server needs to challenge a new user, the first phase marks the user with a rejected ($r$) state. If the user solves the challenge precisely, the state is then marked as successful ($s$). When the user makes a slight mistake, the server marks the user with a pending ($p$) state. From the pending state, only if the user completely solves the next CAPTCHA, the pending state is changed to a successful state; or else, the state retreats to the first phase as the rejected state. Figure \ref{fsm} shows the Finite-State Machine (FSM) diagram of the CAPTCHA generation process.

Here, we let $\alpha$ be the probability of solving a CAPTCHA to be completely correct, $\beta$ be the probability of making a slight mistake, and $\gamma$ represent the probability of giving a wrong response. According to the FSM, the probability of each state is calculated as follows: 
\[\begin{array}{l}
{s_n} = \alpha ({s_{n - 1}} + {p_{n - 1}} + {r_{n - 1}})            \\
{p_n} = \beta ({s_{n - 1}} + {r_{n - 1}}) \\
{r_n} = \beta {p_{n - 1}} + \gamma ({s_{n - 1}} + {p_{n - 1}} + {r_{n - 1}})\\
\alpha  + \beta  + \gamma  = 1
\end{array}\]
where ${s_0} = 0$, ${p_0} = 0$ and ${r_0} = 0$.


\section{Evaluation and analysis} 
\subsection{Implementation}
All algorithms involved in the experiment are implemented using Node.js and MongoDB in the server-side. The front-end webpage development utilizes the React-Bootstrap with the Webpack framework and a collection of hypertext markup language (HTML). All the participants in the study are asked to try three types of the CAPTCHAs which are the character-based case, datagram-based case and the common text-based CAPTCHA as the baseline. We record the participants' reactions to the challenge and examine two performance metrics, response time and success rate, five times for each type. 

To recreate the general conditions, they finish the tests using their own devices ranging from computers, mobiles, laptops to tablets (see Figure \ref{impl3}). And they use a mix of web browsers to access the experiment server. More than 270 volunteers take part in our user studies. Experiment subjects include teachers, students, and the employees of three colleges. Their ages range from 18 to 65, and their education levels range from high school diploma to Ph.D. 
\subsection{Usability Study}
In this study, 
We evaluate the usability of our principle based on three aspects which are partially quoted from \cite{Nielsen03}. 

\subsubsection{Learnability and Efficiency}
\begin{figure}[!t]
\centering
\subfloat{\includegraphics[height=1.6in]{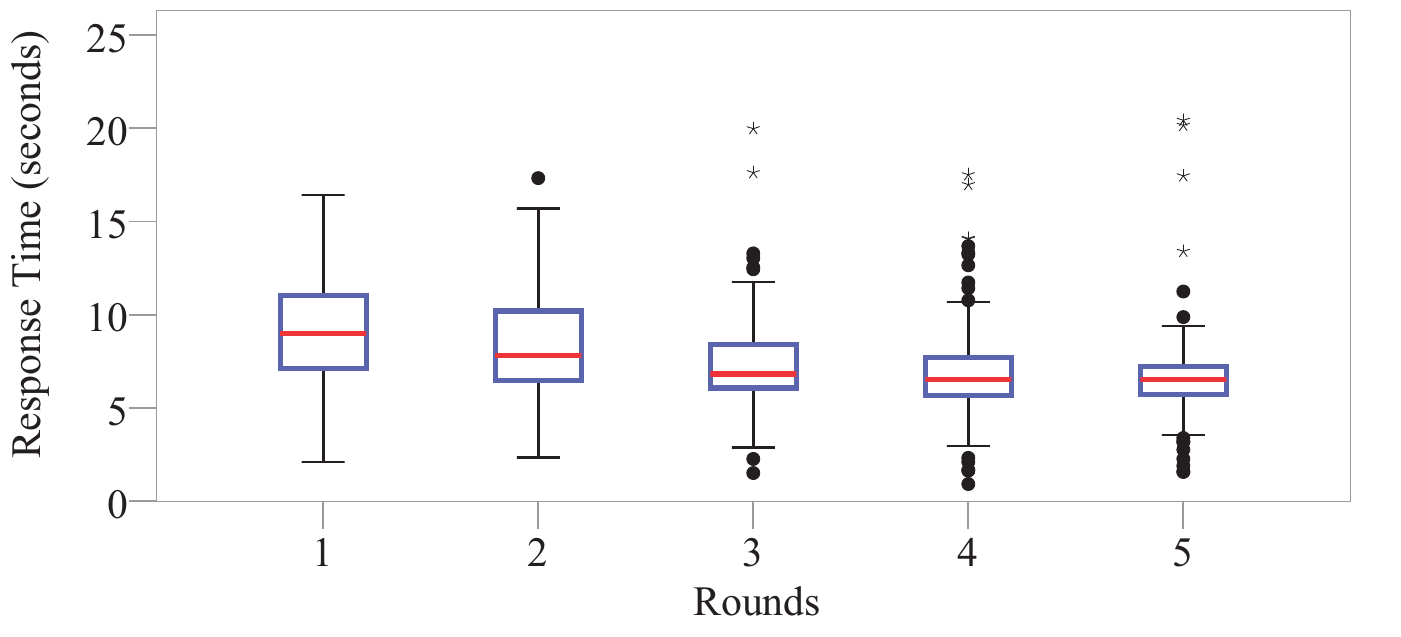}}
\hfil
\subfloat{\includegraphics[height=1.6in]{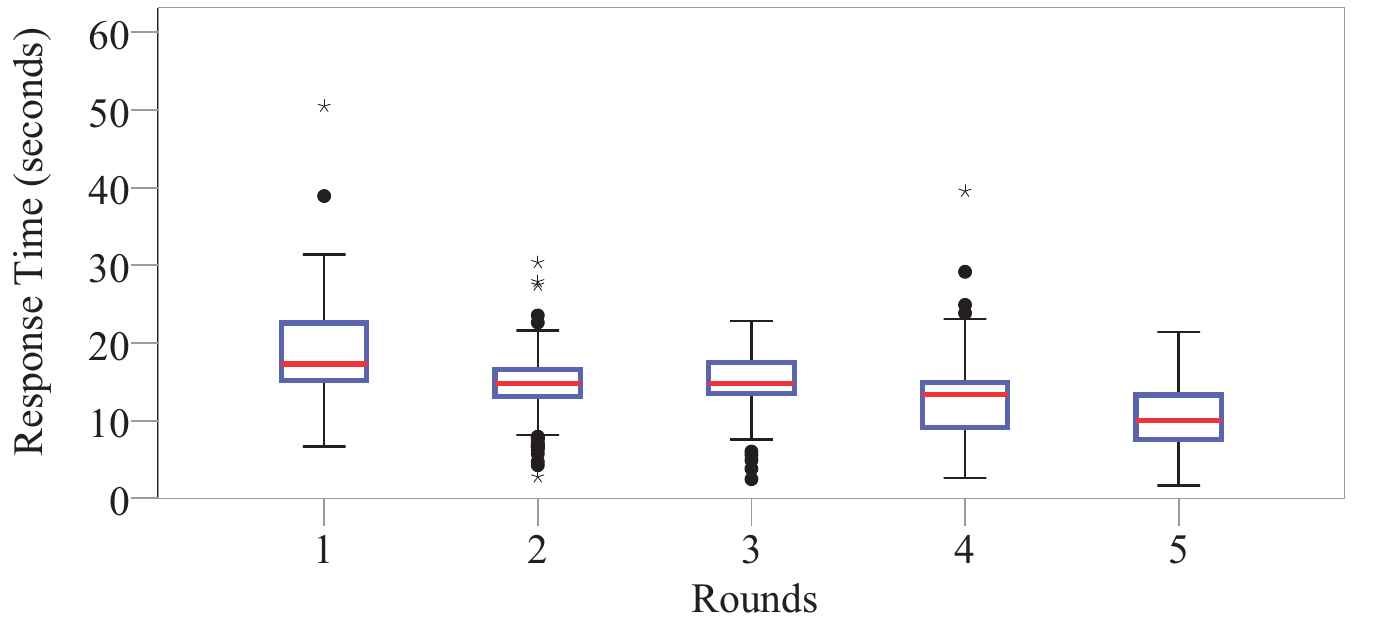}}
\caption{Distribution of character-based (left) and datagram-based (right) challenge response times for each rounds}
\label{timeCI+CII}
\end{figure}

\begin{figure}[!t]
\centering
\includegraphics[height=1.8in]{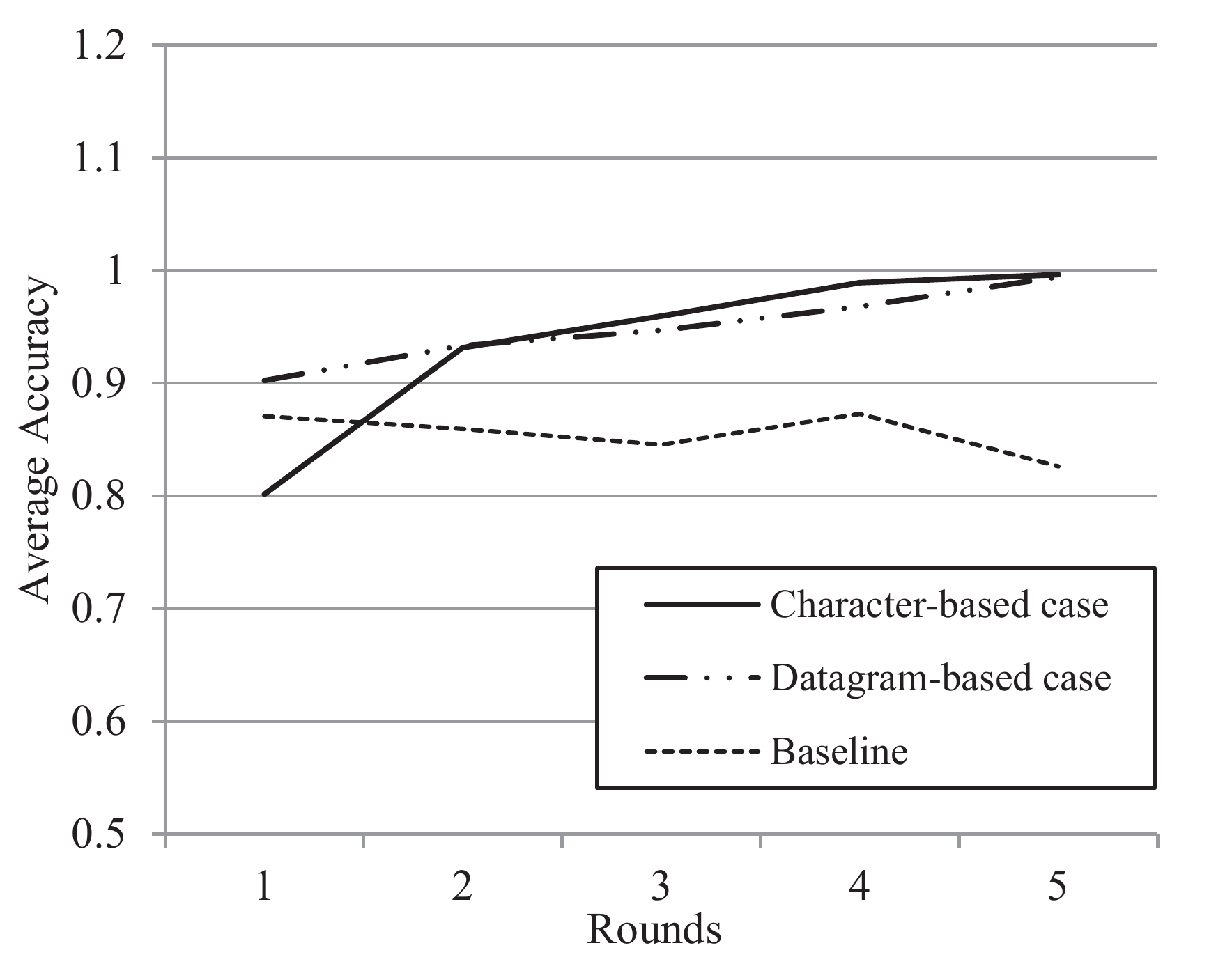}
\caption{The average accuracy of the two cases and baseline in each round}
\label{acc}
\end{figure}

As can be seen in Figure \ref{timeCI+CII}, the user spends more time in the first round. The datagram-based case takes longer than other tests to finish for the volunteers. It costs less than 20s on average to solve the first challenge. However, surprisingly, it is the highest accuracy test with 90\% success rate in the first round according to the Figure \ref{acc}. As the participants' experiences built up, we can find that the response time is reduced quickly from 20s to 10s or below after five rounds. 



In Table \ref{total}, it shows the overall time and the success rate to finish the different kinds of CAPTCHA schemes. We see that there's no significant difference in the accuracy. Typing text takes longer than click action. So our method spends little time than others with similar accuracy. According to these results, both the response time and the success rate suggest that our method is easy to learn, simple to use, and inexpensive to implement. 

%


\begin{table}[tbp]
\centering
\caption{The average response time and accuracy of recognition}
\label{total}
\begin{tabular}{@{}llcc@{}}
\toprule
                                & {CAPTCHA Scheme} & Times(s) & Accuracy(\%) \\ \midrule
\multirow{3}{*}{Our Experiment  } & Character-based Example            & 7.51     & 93.6         \\
								& Datagram-based Example            & 14.29     & 94.9         \\
                                & Baseline                           & 12.72    & 85.5         \\ \midrule
\multirow{4}{*}{Uzun et al.\cite{rtcapt}}    & reCaptcha(\textit{num})                     & 22.11    & 96.7         \\
                                & reCaptcha(\textit{phrase})                  & 20.88    & 91.5         \\
                                & Ebay                               & 12.33    & 100          \\
                                & Yandex                             & 15.05    & 96.7         \\ \bottomrule
\end{tabular}
\end{table}

\subsubsection{Errors and Recovery}
After examining each challenge and the corresponding response, we find that there are three main reasons for these errors: Some of them were attributable to the confusion of the CAPTCHA characters. With the limitation of the fonts that provided by the operation system, many characters, i.e., the number 0 and character O, are not easy to recognize. 
Though rare, some errors occur because some old participants are not proficient enough with standard input devices to interact with the CAPTCHA. When deploying the usability testing, these old people's mouse movements usually make mistakes such as performing a drag action during a single-click. 


\subsection{Security Study}
\subsubsection{Private Data Breach} 
Users may think there will be a risk of password leakage. According to our method, the real password is only used in the user authentication interactions and does not appear at any step of the Turing test. 

For the server-side, it can still hash passwords immediately to limit the exposure of plaintext passwords to threats. All the challenge literal data are coming from the user's POST requests or submissions rather than from the servers' databases. Thus the password strength is not weakened. Besides, for the user-side, 
the attacks may infect the process of the CAPTCHA generation by the authentication data that attackers submit. These fabricated data obtained by performing a CAPTCHA test, hence, are not the real credentials. The adversaries cannot gain access to the web service. 

\subsubsection{Brute Force, Random Guess and Shoulder-Surfing Attack} 
When received a wrong CAPTCHA response, the web server will generate a new challenge image. Therefore, the brute force attack on the CAPTCHA is invalid under our proposed principle. The effectiveness of the random guess attack is mainly dependent on the success rate of the attack. According to the \cite{chellapilla2005designing}, a standard acceptable success rate should not be higher than 0.01\%. Moreover, with the token bucket scheme proposed in \cite{elson2007asirra}, the tolerable success rate can be relaxed to 0.6\% (\emph{TB-Refill} is 3). If we ask users to select the right answer from the eight candidate characters or symbols in the user study and if the adversaries have no basis for selecting the correct characters, the success rate of this attack is probably $1/2^8$ (1/256 or less than 0.4\%). 

Shoulder-surfing is a type of attack to obtain the private information, such as passwords, by looking over the victim's shoulder. Most password filling screens use asterisks instead of showing characters in clear-text. If parts of the password are shown through the CAPTCHA, it may introduce a leakage of private information. 
In fact, it is not so severe as is generally assumed. First, many companies and services 
now require users to meet stringent requirements for password strength and complexity
. It reduces the duplication of letters in the password and makes more difficult to guess it by shoulder-surfing. Second, the asterisk is mainly used to hide the context of characters rather than the characters themselves. The private information contained in the CAPTCHA is selected randomly. It will not reveal too much context information about the password under the circumstances of incorrect responses. Even under the circumstance with the right CAPTCHA selections, purely by shoulder-surfing, the adversaries cannot learn more about the information such as the length of the password or the order of these characters.

\section{Discussion}
\subsection{Application Scope.} The principle we proposed, simple but generic, can work in most web application scenarios. If the information comes from users, it is clear that the proposed approach is suitable for the case when using CAPTCHA for authentication or the authenticated users want to solve CAPTCHA tasks automatically (e.g., intended to crawl some data from a site, etc.). 

\begin{figure}[t]
\centering
\includegraphics[height=1.5in]{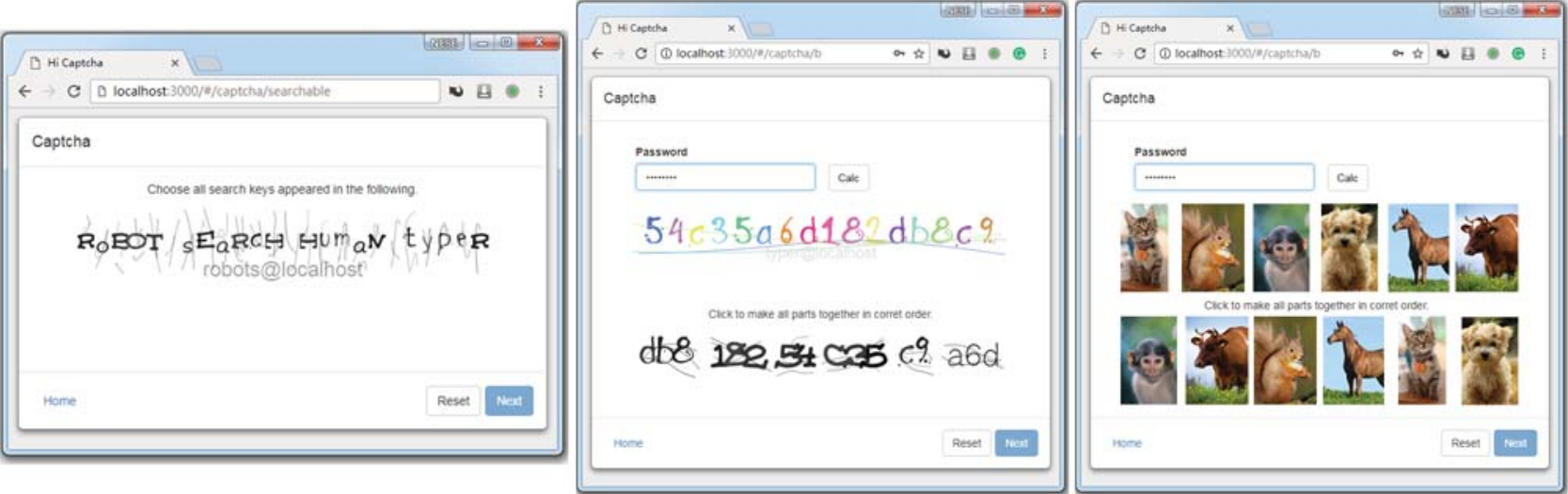}
\caption{The left is an example used in the non-authentication scenarios (e.g., search engine).The middle is the calculation version. The right shows an example of mapping version.}
\label{CIIBCsearch}
\end{figure}

Besides, the CAPTCHA is widely used for non-authenticated applications like avoiding crawlers or abuse of search engine. A striking example is as follows: when sending searches from a robot, search scraper or using software that sends searches to Google, the attacker might be challenged by a reCAPTCHA \cite{von2008recaptcha}. In this circumstance, there is no authentication information shared between client and server. Following our principle, we can randomly select search histories or pages that the user has visited as the user information and embed them into the generation process of the CAPTCHAs. Figure \ref{CIIBCsearch} (left) illustrates the example used in a search engine. If the robot (or attacker) queries the CAPTCHA challenges, the Typers will have no idea for the words meaning unless robot packages all the search histories and sends to them. 

However, we note that the selection of such information is very challenging. Moving forward, we should find more suitable information or a proper expression (encryption) method according to the specific circumstances. 
Without concerning the privacy, we can utilize the asymmetry of information costs or information entropy. This is beyond the discussion scope of this paper. We will do more research on how to solve the Typer problem for the non-authenticated situations in future.

\subsection{Machine Vision Attacks}
The two examples of our case study
are based on the hypothesis of the robustness of the image CAPTCHA. 
If the breakthrough of the recent advancements in automatically solving computer vision tasks (especially the deep learning era) leads an increasing number of CAPTCHAs to become ineffective, our method will face the same problem. 
Following our principle, the characters in the CAPTCHAs are context dependent. For Typers, even if they automate the recognition of the CAPTCHAs, without the private information, it is also unfeasible to pass the test. 

However, the Figure \ref{impl3} (left bottom) shows that case II may be easily solved by reasoning on the order of word segments or be easily attacked by matching features (e.g., colors) of letters in the different segments. 
In a more realistic scenario, the server will not offer the reference object (the hash value) to the user. More specifically, users submit their username and hashed-password to the webserver. Once the server receives the sequences, It will generate the CAPTCHAs by messing up the order. Figure \ref{CIIBCsearch} (middle) shows the CAPTCHAs challenge page. The webserver does not provide the hashed-password but the hash function. And users calculate the reference object themselves. 

\subsection{Password, Graphical Password and CAPTCHA}
Graphical Password is first introduced by Greg Blonder \cite{blonder1996graphical}, may be a solution to the Typer problem to some degree. The idea 
is to let the user click on several regions in an image that appears on the screen. To log in, the user has to click on the same areas again. We can think of this solution as a password-pad of the e-banking system. CAPTCHA as Graphical Passwords (CaRP), proposed by Zhu \emph{et al.} \cite{zhu2014captcha}, is a particular case of the clicking-based graphical password. Hence, it functions as both the graphical password and the CAPTCHA. 

Some people may think separating one human from another is done via passwords. However, passwords and CAPTCHAs have entirely different design purposes. 
By combining the graphical password and the CAPTCHA into a single entity, CaRP changes the login application flow and the basic authentication infrastructure. The application flexibility, password strength, and the system security performance are questionable in its case, too. 

On the other hand, CaRP can be seen as a kind of a particular case of password mapping. The datagram contains the information that defines the relationships between the user clicks and the passwords that the verification process needs. Like our proposed principle, CaRP introduces the password, as the shared private information, into the challenge process. The password can be recovered from the received coordinates which are packaged in the datagram. We think of this process as a password mapping. 
But our approach does not modify the existing authentication framework nor does it reduce the strength of the password, and we can expand the range of applications to solve the non-authenticated situation. 

\subsection{Guide for the Future CAPTCHA Design}
The operation of the datagram-based example is very complicated and hard to use. We should map the data to a proper representation space. 
Here we give an example of a friendly mapping approach to the web interface design industry. As shown in the Figure \ref{CIIBCsearch} (right), the webserver separates the hash into six parts and maps them as six different random animals. The users calculate their private information to get the correct reference object and 
sort cute animals into the correct order. 

The most suitable information to share should be selective, and relatively simple and easy to maintain. For example, we can use the SSNs (Social Security numbers), user passwords or password hashes as this alternative knowledge. 
The attackers will not share the password or other login information with Typers to avoid them stealing these accounts. So we select the password (or hash) in the methodology as the example to demonstrate our principle. Following this principle, We neither store the data (the user uses to verify their password) submitted by the user nor reduce the security of the authentication scheme. The password (or hash) should be the first option in a web application. 

\section{Conclusions}
In this paper, we presented a new CAPTCHA design principle to distinguish human (Typer) from human (user). We first pursue a comprehensive analysis of the Typer phenomenon and the attack mechanism of CAPTCHA. Then we propose our design principle. That is we should leverage the information asymmetry to break the same verification connection for the malicious industry and the Typer. The intuition behind our proposed principle is that different users should share different information with the server and the challenge processes should include this information.

We then formalize, design and implement an example on our principle. A novel generation algorithm for CAPTCHA has been proposed to add the capability of human error tolerance and the difficulty of random guess attack. After the evaluation, both the usability and the security study show a close, if not better, result than the state-of-the-art of CAPTCHA method. The Typer can only deploy a random guess attack with a very low success rate, while a legitimate user can solve the CAPTCHA by humans accurately within less than 15 seconds. The result demonstrates that our proposal meets both security and usability requirements for a good CAPTCHA design. 

As a framework of CAPTCHA, the two examples are just used to demonstrate how our proposal solves the Typer problems. In the future, we can either reduce the distortions of the characters and the number of alternatives or find a more appropriate mapping method to improve the usability and the user experience. If our examples get cracked, a new scheme may appear following our core idea. By this principle, the CAPTCHA can have such a capacity to distinguish a human from another human by attaching the user's private information to avoid the crowdsourcing human labors working in the field who are cheating on CAPTCHA.

\end{document}